\documentstyle{article}
\newcommand{\be}{\begin{equation}}
\newcommand{\ee}{\end{equation}}
\newcommand{\by}{\begin{eqnarray}}
\newcommand{\ey}{\end{eqnarray}}
\newcommand{\lb}{\left (}
\newcommand{\rb}{\right )}
\newcommand{\g}{\gamma}
\newcommand{\G}{\Gamma}
\newcommand{\al}{\alpha}
\newcommand{\bt}{\beta}
\newcommand{\la}{\lambda}
\newcommand{\si}{\sigma}
\newcommand{\om}{\omega}
\newcommand{\ep}{\epsilon}
\newcommand{\zsl}{z \!\!\!/}
\newcommand{\usl}{u \!\!\!/}
\newcommand{\bm}{\bibitem}
\begin{document}
\rightline{BUTP-97/26}
\begin{center}
{\large \bf Operator product expansion at finite temperature} \\
\vspace{1.3cm}
S. Mallik \footnote {Permanent address : Saha Institute of Nuclear Physics,
1/AF, Bidhannagar, Calcutta - 700064, India.}\\
Institut f\"{u}r theoretische Physik der Universit\"{a}t Bern, Sidlerstrasse 5, 
CH-3012 Bern, Switzerland
\end{center}

\vspace{1.3cm}

\noindent {\bf Abstract} \\
We extend an earlier, configuration space
method to find the Wilson coefficients of operators appearing
in the short distance expansion of {\it thermal} correlation functions of 
different
quark bilinears. Considering all the different correlation functions, there 
arise,
up to dimension four, two new operators, in addition to the two appearing 
already in the {\it vacuum} correlation functions. They would contribute  
substantially to the QCD sum rules, when the temperature is not too low.

\vspace{1cm}

The operator product expansion of two currents at short distance  \cite{Wilson}     
played a key role in the past in writing down the extensively used QCD sum rules   
\cite{SVZ}. The vacuum matrix elements of the local operators appearing
in the expansion were used to parametrize the nonperturbative effects of the QCD 
theory at low energy. Using dispersion representations for the correlation 
functions,
these sum rules relate the multitude of low energy physical 
quantities, like the masses and coupling constants of hadrons, to these 
parameters of 
theoretical interest.

In extending these sum rules to finite temperature, two sources of complications 
arise. One is the interaction of the current with the particles
of the medium (heat bath), modifying the dispersion representation \cite{Boch}.
The other is the breakdown of Lorentz invariance by the choice of the reference
frame, in which the heat bath is at rest at a definite temperature. The residual
O(3) invariance allows, in general, more operators of the same dimension in the
operator product expansion than at zero temperature \cite{Shur}.

While the effect of the medium was included in the very first paper \cite{Boch}
on thermal QCD sum rules, the non-perturbative contributions due to the 
additional operators were seriously considered much later \cite{Hatsuda}.
So far only the thermal sum rule for the vector current has been written
with these contributions and analysed at low temperature \cite{Eletsky}.
It still remains to explore the consequences of various sum rules for 
correlation functions of different quark bilinears. One ingredient 
necessary to write them is the set of coefficients with which the operators,
at least the leading ones, contribute to the different sum rules.

Earlier we employed a configuration space method \cite{Fritzsch} to calculate 
the 
coefficients of gluon operators up to dimension six, which appear in the  
vacuum expectation value of correlation functions of 
quark bilinears \cite{Hub},\cite{Smilga}.
It is based on finding the singular solution for the quark propagator  
in the background field of the gluons. It finds a natural 
extension to thermal correlation functions. Here we find the coefficients 
of all operators up to dimension four, appearing in all correlation functions 
built out of any two quark bilinears. 
We start with a brief review of this method.
 
Let $\psi (z)$ be the quark field of a single flavour of mass $m$, where its 
SU(3)
colour indices are suppressed. Its equation of motion is
\be
(-iD \!\!\!\!/ +m)\psi(z) = 0, \qquad  
D_{\mu} = \partial _{\mu}-iA_{\mu},
\qquad A_{\mu}=g A^i_{\mu} {\lambda}^i /2 ,
\ee
where $A_{\mu}(z)$ is the colour gauge field and ${\lambda}^i$ are
 the SU(3) Gell-Mann matrices. 
Assume, to begin with, that  
$A_{\mu}$ is a given classical field. Then the time ordered product 
of the quark fields may be split  as,
\be
T\psi_a (z) \bar{\psi}_b (y) = -iS_{ab}(z,y) + : \psi_a (z) \bar{\psi}_b (y):, 
\ee
where $a, b$ are the spinor indices. The first term is a c-number function,
containing the short distance singularities of the time ordered
product and satisfies  
\be
(-iD \!\!\!\!/_z +m) S(z,y) = \delta^4 (z-y). 
\ee
The second term is a bilocal operator with finite matrix elements as 
$z\rightarrow y$ 
and involves quark dynamics at large distances.

We now outline the method of solving  eq.(3). We  set\footnote{ Operators 
linear in $m$ can be easily sorted out later.}  $m=0$. It is very convenient to 
choose
the Fock-Schwinger gauge \cite{Sch}, $(z-y)^{\mu} A_{\mu} (z) =0$, where the 
gauge
potential $A_{\mu}(z)$ can be expanded in terms of the field strength and its 
covariant derivatives at $z=y$. In the following we put $y=0$ and get,
\[A_{\mu}(z) ={1\over 2} z^{\al} G_{\al \mu}(0) + \cdots .\]
Then the solution will automatically be gauge covariant. 
We convert (3) into a second order differential equation by setting  
\be
S(z,0)= i D \!\!\!\!/ E (z, 0).
\ee
Then $E(z,0)$ satisfies
\be
(\Box -P_{\mu} \partial_{\mu} -Q) E(z,0) =\delta^4 (z) ,
\ee 
where
\[P_{\mu}(z)  =i z^{\al}G_{\al\mu} +\cdots,  \qquad Q(z) = {i\over 2}
 \g G \g + \cdots .\]
Here and below we use matrix notation,  eg., $\g G\g =\g^{\mu} G_{\mu \nu} 
\g^{\nu}$. 
The fundamental or the singular solution to (5) is now given by
the Hadamard Ansatz for $E(z,0)$\footnote{For the time ordered propagator
$z^2$ stands for $z^2 -i\ep$} \cite{Courant},
\be
E(z,0) =-{i\over {4\pi^2}} \lb {1\over z^2} + V(z) ln(4z^2 {\mu}^2) + W(z) \rb ,
\ee
where $V(z)$ and $W(z)$ are analytic functions of $z$, regular in the 
neighbourhood of $z=0$. Although not singular, $W(z)$ is added to ensure that
$E(z, 0)$, by itself, satisfies eq.(5). The factor $4\mu^2$ in the argument of 
the 
logarithm represents a mass scale  (not related to the quark mass),
arbitrary at this stage.
Inserting this Ansatz in (5) we get the equations 
for $V$ and $W$, which can be readily solved in  power series in $z_{\mu}$.

The solution for $S(z,0) $ is worked out in ref.(8) up to gauge field strengths 
of
dimension six. Here we reproduce this solution up to dimension four,
\by
S(z,0) &=& {1\over {4\pi^2}}  \Biggl[ \partial \!\!\!/ ({1\over z^2}) -{i\over 
{4z^2}} 
\g^{\mu} z \!\!\!/ \g^{\nu} G_{\mu \nu} +{1\over {24z^2}} z \!\!\!/ (zG\g)^2  
\Biggr. 
\nonumber \\
       & &+\lb -{1\over 24} zGG\g + {1\over 16} (zG\g)(\g G \g) +{z \!\!\!/ 
\over 32}
\{ {1\over 3} \g GG\g -{1\over 2}(\g G \g)^2\}\rb ln (4z^2 \mu^2)  \nonumber   
\\
       & & \Biggl. -{z \!\!\!/ \over 64} \{ {1\over 3} \g GG\g -{1\over 2}(\g G 
\g)^2\} \Biggr]   
\ey
Again the last term, which is regular, ensures that $S(z,0)$, by itself,  
satisfies $(3)$
up to dimension four. Here we have omitted terms 
involving the gauge field covariant derivatives, as they do not survive the 
trace
over colour in the correlation functions. In this case $S(z,0)$ satisfies,
$S(0,z)=S(-z,0)$   

Once we have found the c-number coefficients multiplying the classical gauge 
field
strengths in $S(z,0)$, the latter can be considered as gauge field operators of 
QCD theory.
Eqs. (2) and (7) thus constitute the short distance expansion of the two point 
function of the quark fields. Clearly, higher order corrections to the 
coefficients
due to gluon propagation is beyond the present method.

Consider now the short distance expansion of the 
time ordered product of two quark bilinear operators,
\[ J_1 (z)= :\bar{\psi} (z) \G_1 \psi(z): , \qquad  J_2(z) = :\bar{\psi} (z)\G_2 
\psi(z): ,\]
where $\G_1$ and $\G_2$ are any two of the Dirac matrices. Using (2) it can be 
written
as a sum of three types of terms,
\[ TJ_1(z) J_2(0) = t^{(G)}_{12} +t^{(\psi)}_{12} +t^{(reg)}_{12}, \]
where
\by
t^{(G)}_{12} &=& tr \G_1 S(z,0)\G_2 S(0,z) , \\
t^{(\psi)}_{12} &=& -i(\G_2 S_m (0,z) \G_1)_{ab} :\bar{\psi}_a (0) \psi _b (z): 
        -i(\G_1 S_m (z,0) \G_2)_{ab} :  \bar{\psi}_a (z) \psi_b (0) : , \\
t^{(reg)}_{12} &= & :\bar{\psi} (z) \G_1 \psi (z) \bar{\psi} (0) \G_2 \psi (0): 
,
\ey
where tr indicates trace over both Dirac and colour matrices.
In (9) we have included the missing term in $S(z,0)$, which is linear in the 
quark mass,
\[ S_m(z,0) = S(z,0) - \frac{im}{4\pi^2z^2}  . \]
The first term gives the coefficients of the gluon operators. The second term, 
after expanding
$\psi (z)$ in the bilocal operators as,
\[ \psi (z) = \psi (0) + z^{\mu} D_{\mu} \psi(0) +\cdots, \]
gives the coefficients of all quark operators,
\by
t^{(\psi)}_{12} &=& -{1\over {2\pi^2 (z^2)^2}} \Bigl[ \left \{ i(\G_2 \zsl \G_1
-\G_1 \zsl \G_2)_{ab} +mz^2 (\G_1\G_2  + \G_2\G_1)_{ab}/2\right \} :\bar{\psi}_a 
\psi_b:
  \Bigr. \nonumber \\
  & & +(\G_2 \zsl \G_1 + \G_1 \zsl \G_2)_{ab}  z^{\mu} :\bar{\psi}_a i D_{\mu} 
\psi_b:
 \Bigr]
\ey
The third term is of no interest to us.

Thus $t^{(G)}_{12}$ and $t^{(\psi)}_{12}$ constitute the most general operator 
product 
expansion of two quark bilinears,
giving the coefficients of all the gluon and quark 
operators up to dimension four, having various tensor and spinor structures.
Clearly at the level of operators there is no reference to temperature.
It is in taking the matrix elements of these operators that the difference
between the vacuum state and the thermal state shows up. In taking the vacuum 
expectation 
value, the different tensor and the spinor operators are projected on to their 
corresponding Lorentz scalars. The situation is
different for the thermal expectation value, which for any operator $O$ is 
defined as
\[ <O>_T = Tr e^{-\bt H} O/ Tr e^{-\bt H}, \]
where $H$ is the QCD Hamiltonian,  $\bt$ is the inverse temperature T and 
the Tr(ace) is over any complete set of states. Here the choice of the thermal 
rest frame, where the temperature is defined, breaks the Lorentz invariance
and we have to include $O(3)$ invariant operators as well.

It is, however, convenient to restore Lorentz invariance formally in the thermal 
field theory by introducing the four-vector $u^{\mu}$, the velocity of the heat
bath. [ $ u^2=1 $ and $u^{\mu}= (1,0,0,0) $ in the rest frame of the heat bath.]
Thus while at zero temperature scalar operators are formed from tensors 
by contracting its indices among themselves, at finite temperature additional 
scalars 
can be formed by contracting its indices with $u^{\mu}$.
  
Consider first the piece $t^{(G)}_{12}$. The Lorentz covariance at finite 
temperature now allows us to write
\by
<tr^c G_{\al \bt} G_{\la \si}>_T  &=& (g_{\al \la} g_{\bt \si} -g_{\al \si} 
g_{\bt \la})A \nonumber \\
                         & & \mbox{} -(u_{\al} u_{\la} g_{\bt \si} -u_{\al} 
u_{\si} g_{\bt \la}
                             -u_{\bt} u_{\la} g_{\al \si} +u_{\bt} u_{\si} 
g_{\al \la})B .
\ey                             
By contracting indices on both sides, we get
\by
A &=& {1\over 24} <G^a_{\al \bt} G^{a \al \bt}>_T 
+{1\over 6}<u^{\al} {\Theta}^g_{\al \bt} u^{\bt}>_T, \\
B &=& {1\over 3}<u^{\al} {\Theta}^g _{\al \bt} u^{\bt}>_T,
\ey
${\Theta}^g_{\al \bt}$ being the traceless,
 gluonic part of the stress-tensor of the QCD theory,
\be
{\Theta}^g_{\al \bt} =-G^a_{\al \la} G_\bt^{\la a} 
+{1\over 4} g_{\al \bt} G^a_{\la \si} G^{\la \si a} .
\ee
Since we are interested in the coefficients of operators of dimension four only, 
we may 
already use (12) in the expression for $S(z,0)$, which then simplifies to
\be
S(z,0)={1\over 4\pi^2} \lb \partial \!\!\!/({1\over z^2}) -{i\over 4z^2}
\g^{\mu} \zsl \g^{\nu} G_{\mu \nu} + {B\over 48} (a\zsl +b\usl) \rb ,
\ee
where

\[ a = 1-{4\over z^2} (u\cdot z)^2-2ln(4z^2 \mu^2),  \qquad
b = 8u\cdot z ln(4 z^2 \mu^2) . \]
Insert this expression for $S(z,0)$ in $t^{(G)}_{12}$ and use (12)  again for 
the 
dimension four operators generated in the product. Then it only remains to 
evaluate the
traces over $\g$-matrices for different correlation functions (specified by 
$\G_1$ and $\G_2$ ) to 
find the coefficients of the operators in A and B. 

Next we turn to the piece $t^{(\psi)}_{12}$. As with the gluon operators,
we project the quark operators
on to the corresponding scalars.  
At finite temperature the most general decomposition of the two operators in 
(11) 
in spinor space, conserving parity, are given by         
\by
<\bar{\psi}_a \psi_b>_T  &=& {\delta_{ab}\over 4}<\bar{\psi} \psi>_T 
+{(\usl)_{ba}\over 4}
<\bar{\psi} \usl \psi>_T  \\
<\bar{\psi}_a iD_{\mu} \psi_b>_T &=& (\g_{\mu})_{ba}U +u_{\mu} (\usl)_{ba} V 
+u_{\mu} \delta_{ba} Y +(\usl \g_{\mu})_{ba} Z
\ey
where
\be
U={m\over 16}<\bar{\psi} \psi>_T -{1\over 12}<u^{\mu} \Theta^f_{\mu 
\nu}u^{\nu}>_T,\qquad
V={1\over 3}<u^{\mu} \Theta^f_{\mu \nu} u^{\nu}>_T ,
\ee
\be
Y={1\over 6} <\bar{\psi} iu\cdot D \psi>_T +{m\over 12}<\bar{\psi} \usl \psi>_T, 
\qquad
Z={1\over 12}<\bar{\psi}iu\cdot D\psi>_T -{m\over 12}<\bar{\psi}\usl \psi>_T
\ee
and $\Theta^f_{\mu\nu}$ is the traceless, fermionic part of the stress tensor,
\[ \Theta^f_{\mu\nu} = \bar{\psi} \g_{\mu}i D_{\nu} \psi 
-{1\over 4}g_{\mu \nu} m\bar{\psi} \psi .\]
Now $\bar{\psi}iu\cdot D\psi$ can be reduced to $m\bar{\psi}\usl \psi$ and a 
total derivative.
Further $ <\bar{\psi}\usl \psi>_T = 0 $, for zero chemical potential and we 
shall not 
consider it any further. We are thus left with only $<\bar{\psi}\psi>_T$ in (17) 
and $U$ and $V$ in (18).
Inserting these decompositions in $t^{(\psi)}_{12}$ and evaluating the traces 
over the $\g$-matrices, 
we can get the coefficients of these operators. 
         
We have thus reduced the calculation of the Wilson coefficients of operators 
appearing 
in the expansion of the product of any two 
quark bilinears to evaluating traces over $\g$-matrices.
These coefficients have to be Fourier transformed to momentum space for use in 
the 
QCD sum rules. Here we only state this result for the correlation function of 
vector 
currents $(\G_1=\g_{\mu}, \G_2 =\g_{\nu})$. It can be expressed through two 
invariant 
amplitudes, $T_1$ and $T_2$, say, which are functions of $\om = u\cdot q $ and 
$q^2$,                           
\be
i \int d^4 z e^{iq\cdot z} <T J_{\mu} (z) J_{\nu}(0)>_T 
= \lb -\eta_{\mu \nu} +{q_{\mu} q_{\nu}\over q^2} \rb T_1 (\om, q^2)  + \lb  
u_{\mu} -{\om q_{\mu}
\over q^2} \rb \lb u_{\nu} - {\om q_{\nu}\over q^2} \rb T_2 (\om, q^2)
\ee
Then we find  $(Q^2 = - q^2)$,
\by
T_1&=&{2m\over q^2} <\bar{\psi} \psi>_T + {g^2\over {48 \pi^2 q^2}} <G^a_{\mu 
\nu} G^{\mu \nu a}>_T
-{8\over 3 q^2} \lb 1-{2 \om^2\over q^2} \rb  <u\Theta^f u>_T  \nonumber  \\ 
   & & -{g^2\over {9\pi^2 q^2 }}
    \left \{ \lb 1-{2\om^2\over q^2} \rb \lb ln(Q^2/\mu^2)  +2\g \rb
     -{5\over 4} + \frac{7\om^2}{2q^2}  \right \}
    <u\Theta^g u>_T,   \\   
T_2&=&-\frac{16}{3 q^2} <u\Theta^f u>_T  -\frac{2g^2}{9\pi^2 q^2 } \lb  
ln(Q^2/\mu^2) 
+2\g -{3\over 4} \rb <u\Theta^g u>_T ,
\ey         
where $\g$ is Euler constant.

The large $q^2$ behaviour of the Wilson coefficients can be improved by the 
renormalisation group
equation. It effectively shifts the renormalisation scale of these coefficients 
from $\mu$ to
$\sqrt{Q^2}$. The operators $g^2 G_{\mu \nu} G^{\mu \nu}$ and $m\bar{\psi} \psi$ 
and hence their
coefficients are invariant under this change of scale. But it mixes the 
operators $\Theta^f_{\mu \nu}$
and $\Theta^g_{\mu \nu}$, a situation familiar from the analysis of deep  
inelasic scattering
\cite{Politzer}.
Let us write the contributions of $\Theta^f $ and $\Theta^g$ 
to either $T_1$ or $T_2$ , for short, as
\[ C_f n_f \Theta^f|_{\mu} + C_g \Theta^g|_{\mu} \]
where the coefficients 
$C_{f,g} [\equiv C_{f,g} (log(Q^2/\mu^2), g(\mu^2), \om^2/q^2) ]$  can be 
read off from (22, 23)
and $n_f$ is the number of flavours. The coefficients
satisfy the coupled renormalisation group equations with an anomalous dimension 
matrix \cite{Peskin}.
Diagonalising this matrix, we see that the
combination of the coefficients $(n_f C_f +{16\over 3} C_g)$, corresponding to 
the total
energy-momentum tensor operator $(n_f\Theta^f +\Theta^g)$, has zero anomalous 
dimension, while
the combination $(C_f -C_g)$, corresponding to the operator $n_f({16\over 
3}\Theta^f -\Theta^g)$
has anomalous dimension, $-{g^2\over {(4\pi)^2}} d$, with $d={4\over 3}({16\over 
3} + n_f)$. 
Thus 
\by
& & C_f n_f \Theta^f|_{\mu} + C_g \Theta^g|_{\mu} \nonumber  \\
& & = {1\over {{16\over 3} +n_f}} \left \{ (n_f \bar{C}_f + {16\over 3} 
\bar{C}_g) 
(n_f \Theta^f +\Theta^g) + a n_f(\bar{C}_f -\bar{C}_g) 
({16\over 3} \Theta^f -\Theta^g)|_{\mu} \right \},       
\ey                                  
where $\bar{C}_{f,g} =\bar{C}_{f,g} (1, g^2(Q^2))$ and
$ a= \{(g^2(\mu^2)/g^2(Q^2) \}^{-d/2b}, \quad  b=11-2n_f/3.$
Thus ignoring $g^2$-corrections (which for $C_f$ is not calculable anyway by the 
present method),
the stress tensor contributes to $T_1$, say, as\footnote{Our result does not 
agree with the one stated 
by the authors of the second paper in ref.(6), because they work in the deep 
inelastic limit and do not use the renormalisation group equation.},
\be
T_1 \rightarrow -\frac{8}{(16+3n_f)q^2} (1-2\om^2/q^2) 
 \left \{ \ep +a(16\ep_f/3 -\ep_g)|_{\mu} \right \},
\ee
where we use notation for the rest frame of the heat bath, $\ep =n_f \ep_f 
+\ep_g$, with
$\ep_f$ and $\ep_g$ being the energy densities of quarks and gluons 
respectively.

Finally let us list the set of thermally averaged operators contributing to the 
thermal QCD sum 
rules. Considering all the different correlation functions 
and restricting to operators up to dimension four, there are 
two new operators, namely,  $\ep =(n_f\ep _f+\ep_g) $ and 
 $(16\ep_f /3 -\ep_g)$.  
Two condensates in the vacuum sum rules, now appearing as thermal averages, viz, 
$<\bar{\psi} \psi>_T$ 
and $<G_{\mu \nu}^a G^{\mu\nu a}>_T$, have their temperature dependence given by 
\cite{Leutwyler}, 
\[ <\bar{\psi} \psi>_T = <0|\bar{\psi}\psi|0> -\frac{\partial P}{\partial m}, \]
\[ <\frac{g^2}{4\pi^2} G_{\mu\nu}^a G^{\mu\nu a}>_T = 
<0|\frac{g^2}{4\pi^2} G_{\mu\nu}^a G^{\mu\nu a}|0> -\frac{8}{b} (\ep -3P), \]
where $P$ is the pressure. In the hadronic phase they are known
from chiral perturbation theory \cite{Gerber}, \cite{Bangerter} 
and at high enough temperature from ordinary  QCD perturbation theory. 
Thus only one condensate, viz, $ (16\ep_f/3  -\ep_g) $
cannot be readily evaluated. It may be obtained from one of the
sum rules themselves. It should be noted that though the contributions of the 
new 
operators would be small for small enough temperature, they would, in general, 
dominate
over the condensates in 
the QCD sum rules in the neighbourhood of the phase transition, where the latter 
ones 
begin to 'melt' away.

I have much pleasure to thank H. Leutwyler for many helpful discussions. I also 
thank B.L. Ioffe for bringing 
the references (6) to my attention. I gratefully acknowledge the hospitality 
at the University of Bern, where this work was carried out.

\end{document}